\documentclass[journal]{IEEEtran}

\usepackage{float}

\usepackage{cite}

\ifCLASSINFOpdf
  \usepackage[pdftex]{graphicx}
\else
\fi
\usepackage[table]{xcolor}
\usepackage{multirow}
\usepackage{makecell}

\usepackage{tabularx}
\usepackage{array,tabularx}

\usepackage{array}

\usepackage{algorithm}
\usepackage{algpseudocode}
\usepackage{hhline}

\usepackage{graphicx} 

\usepackage{multirow}
\usepackage{xcolor}
\usepackage{array, xcolor, multirow, makecell}

\usepackage{amsmath}

\begin{document}
%
\title{Optimal Placement and Sizing of PV-Based DG Units in a Distribution Network Considering Loading Capacity}
%

\author{Abhinav~Sharma,~\IEEEmembership{Student Member,~IEEE,}
        Pratyush~Chakraborty,~\IEEEmembership{Senior Member,~IEEE,}\\
        Manoj~Datta,~\IEEEmembership{Senior Member,~IEEE,} and Kazi~N.~Hasan,~\IEEEmembership {Senior Member,~IEEE} 
\thanks{Abhinav Sharma is with the Department
of Electrical and Electronics Engineering, BITS Pilani, Hyderabad Campus, Hyderabad, 500078, India, and also with the School of Engineering, RMIT University, Melbourne
VIC 3000, Australia. }
\thanks{Pratyush Chakraborty is with the Department of Electrical and Electronics
Engineering, BITS Pilani Hyderabad Campus, Hyderabad, 500078, India.}
\thanks{Manoj Datta and Kazi N. Hasan is with the School of Engineering, RMIT University, Melbourne
VIC 3000, Australia.}}
%
%

\markboth{}%
{Shell \MakeLowercase{\textit{et al.}}: }
%



\maketitle

\begin{abstract}
This research paper proposes an efficient methodology for the allocation of multiple photovoltaic (PV)-based distributed generation (DG) units in the radial distribution network (RDN), while considering the loading capacity of the network. The proposed method is structured using a two-stage approach. In the first stage, the additional active power loading capacity of the network and each individual bus is determined using an iterative approach.
This analysis quantifies the network's additional active loadability limits and identifies buses with high active power loading capacity, which are considered candidate nodes for the placement of DG units. Subsequently, in the second stage, the optimal locations and sizes of DG units are determined using the Monte Carlo method, with the objectives of minimizing voltage deviation and reducing active power losses in the network. The methodology is validated on the standard IEEE 33-bus RDN to determine the optimal locations and sizes of DG units. The results demonstrate that the optimal allocation of one, two, and three DG units, achieved from proposed method, reduces network’s active power losses by 50.37\%, 58.62\%, and 65.16\%, respectively, and also significantly enhances the voltage profile across all buses.
When the obtained results are compared with the results of several existing studies, it is found that the proposed method allows for larger DG capacities and maintains better voltage profiles throughout the RDN.

\end{abstract}


             \section*{Nomenclature}
\begin{description}
  \item[$m,n$] Index for buses
  \item[$c$] Index for interconnection point 
  \item[$L$] Set of lines
  \item[$M$] Set of nodes
  \item[$C$] Set of interconnection points 
  \item[$b$] Line susceptance
  \item[$g$] Line conductance
  \item[$P_{m}^{k}$] Additional active power loading
  \item[$P^{M}$] Grid supplied active power
  \item[$Q^{M}$] Grid supplied reactive power
  \item[$PL$]    Line  active power injection
  \item[$QL$] Line reactive power injection 
  \item[$P_D$] Active power load
  \item[$Q_D$] Reactive power load
  \item[$P_{\mathrm{DG}}$] Power output from DG
  \item[$V$] Voltage magnitude
  \item[$\theta$] Voltage angle
\end{description}
%
\IEEEpeerreviewmaketitle
%
%
%
%
\section{Introduction}
\IEEEPARstart{T}{he} increasing penetration of renewable energy sources (RES) into power distribution networks is a critical step toward achieving sustainable and low-carbon energy systems. However, the integration of RES poses significant technical challenges due to the intermittent nature and the limited hosting capacity of existing distribution infrastructure. Traditionally, distribution networks were structured to transmit power unidirectionally from central generating stations to end-users. Accommodating distributed RES requires careful planning to maintain voltage stability, minimize power losses, and enhance network reliability\cite{1},\cite{2}.

In recent years, there has been notable focus on developing efficient techniques for optimal DG allocation in distribution networks. These approaches are mainly intended to improve voltage profiles, minimize losses in the network, and strengthen the reliability and resilience of power system \cite{3}. Numerous optimization strategies have focused on determining optimal allocation of DG units. Broadly, these strategies are divided into three main categories. The first group includes metaheuristic algorithms, such as evolutionary computation and simulated annealing, which are widely adopted for their ability to escape local optima and handle nonlinear search spaces \cite{4}, \cite{5}. The second group involves mathematical programming approaches, where formulations such as linear, nonlinear, and  mixed-integer programming are employed depending on the problem structure \cite{6}, \cite{7}, \cite{8}, \cite{9}. A third category comprises search-based techniques, for example, Tabu search and group search optimization, which rely on systematic exploration of feasible solutions \cite{10}, \cite{11}. Different load flow models are incorporated to evaluate candidate solutions, including backward/forward sweep, probabilistic power flow, and the distFlow formulation \cite{12}, \cite{13}, \cite{14}. Moreover, advanced optimization frameworks like sequential quadratic programming (SQP) and AC optimal power flow (ACOPF) are often used for enhancing solution accuracy and reliability \cite{15}, \cite{16}.

The author of \cite{17} focuses on determining the additional loading capacity to ensure that the system can handle increased demand or integrate distributed generation without compromising network performance. The system's capability to accommodate increased load demand through the integration of wind and solar DG units without violating network constraints is assessed in \cite{18}.

The particle swarm optimization (PSO) algorithm in \cite{19} used for optimal allocation of PV-DG units, aiming to minimize losses, improve voltage deviation, and enhance cost-effectiveness. The joint optimization of RDN reconfiguration with DG allocation has been formulated as MILP problem to overcome the limitations of metaheuristic methods. By linearizing the nonlinear problem, the proposed approach ensures global optimality and convergence\cite{20},\cite{21},\cite{22}.

 A nonlinear programming framework has also been proposed for determining the optimal allocation for renewable DGs, with objective of reducing network losses through localized generation \cite{23}. The approach incorporates ACOPF, while accounting for operational limits and uncertainties in demand and renewable output. Another contribution in this area combines probabilistic nonlinear optimization with sensitivity-based analysis to reduce losses while simultaneously determine DG allocation and transformer tap positions \cite{24}. These methods offered a more effective and reliable alternative compared to traditional planning techniques.

Multi-objective optimization approaches have been adopted to enhance both the loadability limits and reduce losses in distribution networks. One such technique frames the problem of optimal DG allocation as an MINLP model and solves it using a two-stage (bi-level) strategy. The first stage, referred to as the siting planning model (SPM) \cite{25}, utilizes index-based methods to identify promising bus locations. In the second stage, the capacity planning model (CPM) \cite{26}, optimization techniques as SQP and BAB\cite{27} are applied to determine optimal DG sizes. The use of heuristic index-based methods in the siting phase may overlook critical operational constraints, the computational burden of  mixed-integer nonlinear programming-based sizing strategies can hinder practical deployment, especially in larger or real-time applications.

It can be observed from the existing literature that only few studies consider a composite objective function as the minimization of voltage deviation and network active power losses when solving the optimal allocation problem. Most studies focus solely on single optimization criterion, such as reducing power losses or improving the voltage profile. Furthermore, in many existing approaches, vital operational constraints of the distribution system such as line flow limits, equipment ratings, and permissible bus voltage limits are not considered. As a result, the actual capacity of the network to accommodate additional loads or DGs may be overlooked, leading to suboptimal solutions or even infeasible solutions. These limitations have been resolved in this paper by developing an efficient two-stage methodology. The major contributions of this paper are as follows:
\begin{enumerate}
    \item We developed an efficient two-stage methodology for the optimal placement and sizing of multiple DG units in a distribution network. A new advantageous feature of this method is that it first evaluates the loading limits of the network and of each individual bus, ensuring that the DG units have larger capacities and are placed only at nodes with sufficient installation capacity.
    
    \item Our method addresses the complex and highly constrained DG allocation problem by considering all the vital operational constraints of the network.
    
    
    
    \item We validated the method on the standard IEEE 33-bus RDN, demonstrating its effectiveness in significantly reducing active power losses and improving the voltage profile.
    
    \item We conducted comparative analysis of the results obtained from our method with those of several existing approaches, and our method demonstrated better performance.
\end{enumerate}

The structure of this paper is as follows: Section I provides an introduction, including a comprehensive review of related literature, the motivation behind this study, key contributions, and the paper structure. The mathematical modeling is provided in Section II. The proposed methodology is discussed in detail in Section III. The validation of the methodology using the IEEE 33-bus test system is detailed in Section IV, while Section V presents the conclusions of the study.

\hfill
\section{Mathematical Modeling}
In this section, the optimal allocation problem in a distribution network is formulated as a two-stage approach. The mathematical formulation of the first stage, for determining the network’s additional active loadability limits, is modeled, and the mathematical formulation of the second stage, to determine the optimal locations and sizes of DG units using the Monte Carlo method is also modeled. The mathematical modeling for both stages is given below.
\hfill
\subsection{Mathematical modeling of the first stage }
The constrained optimization problem for the first stage is formulated as:\\

\noindent 
\textit{1) Objective function}

The objective is to find the additional active loading capacity, which is given as:

\begin{equation}
\max \sum_{m \in \mathcal{M}} P_{m\lambda}
\end{equation}
\textit{2) Constraints} 

Nodal active and reactive power balances are modeled by equations (2) and (3). They ensure that the total power, which includes both grid supplied power and power injected through connected lines, at each bus equals the corresponding power demand.
\begin{equation}
\sum_{c \in \mathcal{C}_m} P^M_c + \sum_{l \in \mathcal{L}_{nm}} PL_{nm} = P_{dm} + P_{m\lambda} \qquad \forall m \in \mathcal{M}
\end{equation}

\begin{equation}
\sum_{c \in \mathcal{C}_m} Q^M_c + \sum_{l \in \mathcal{L}_{nm}} QL_{nm} = Q_{dm} \qquad \forall        m \in \mathcal{M}
\end{equation}

The power supplied from the slack bus is constrained by specified power limits, defined in equations (4) and (5). 

\begin{equation}
P_c^{M,\min} \leq P^M_c \leq P_c^{M,\max} \qquad \forall c \in \mathcal{C}_m
\end{equation}

\begin{equation}
Q_c^{M,\min} \leq Q^M_c \leq Q_c^{M,\max} \qquad \forall c \in \mathcal{C}_m
\end{equation}

Set $\mathcal{C}_m$ comprises all buses that serve as POIs to the upstream network.
Line flows are evaluated using equations (6) and (7), considering line parameters such as conductance and susceptance, voltage magnitudes, and phase angles at the corresponding buses.
\begin{equation}
\scalebox{0.93}{$
    PL_{mn} = g_{mn} V_m^2 - g_{mn} V_m V_n \cos\theta_{mn}
             - b_{mn} V_m V_n \sin\theta_{mn}
$}
\end{equation}

\begin{equation}
\scalebox{0.905}{$
QL_{mn} = -b_{mn} V_m^2 - b_{mn} V_m V_n \cos\theta_{mn}  - g_{mn} V_m V_n \sin\theta_{mn} 
$}
\end{equation}

The line flows are constrained with their respective power flow limits, as defined in equations (8) and (9).
\begin{equation}
-PL_{mn}^{\max} \leq PL_{mn} \leq PL_{mn}^{\max} \qquad \forall mn \in \mathcal{L}
\end{equation}

\begin{equation}
-QL_{mn}^{\max} \leq QL_{mn} \leq QL_{mn}^{\max} \qquad \forall mn \in \mathcal{L}
\end{equation}

 The voltage magnitude at each bus is constrained within specified upper and lower bounds, as given in equation (10).

\begin{equation}
V_m^{\min} \leq V_m \leq V_m^{\max} \qquad \forall m \in \mathcal{M}
\end{equation}

The maximum current carrying capacity for each line within the corresponding thermal limits is specified as (11).

\begin{equation}
I_{mn} \leq I_{mn}^{\max} \qquad \forall mn \in \mathcal{M}
\end{equation}
\subsection{Mathematical modeling of the second stage }
The constrained optimization problem for the second stage is
formulated as:

\noindent 
\textit{1) Objective function }

Two single objective functions are considered as minimization of voltage deviation and minimization of active power loss.\\
\textit{a) Minimization of Voltage Deviation} 

Voltage deviation ($\Delta V_m$),
\begin{equation}
\Delta V_m = |V_m - 1|,
\end{equation}
\begin{equation}
f_{1} = \min \sum_{m \in \mathcal{M}} |V_m - 1|
\end{equation}
where $V_m$ is voltage profile on $m$th bus.The $\Delta V_m$ represents the deviation between the voltage at the $m$th bus and the reference voltage (1 pu).

\textit{b) Minimization of active power losses} 

\text{Total active power losses} ($P_L$),
\begin{equation}
P_L = \sum_{mn=1}^{L} \frac{PL_{mn}^2 + QL_{mn}^2}{V_m^2} R_{mn},
\end{equation}

\begin{equation}
f_{2} = \min (P_L)
\end{equation}

The weighted sum approach combines multiple single objective functions into a composite function. The overall objective function is formulated by using multiple single objectives and given as:
\begin{equation}
\text{$F_{obj.}$} = (w_1 \times f_1) + (w_2 \times f_2)
\end{equation}

Each objective function is associated with a weight that reflects its relative priority. These weighting coefficients are strictly positive and normalized such that,
\begin{equation}
\sum_{i=1}^{2} w_i = 1 \quad \text{………} \quad w_i \in (0, 1)  \end{equation}
\text{where $w$ is the weighting 
factor.} \\
\hfill

\textit{2) Constraints }

Nodal active and reactive power balances are modeled by equations (18) and (19). They ensure that the total power, which includes both grid supplied power and power injected through connected lines, at each bus equals the corresponding power demand.
\begin{equation}
\sum_{c \in \mathcal{C}_m} P^M_c + \sum_{l \in \mathcal{L}_{nm}} P_{Lnm} = P_{dm}  \qquad \forall m \in \mathcal{M}
\end{equation}
\begin{equation}
\sum_{c \in \mathcal{C}_m} Q^M_c + \sum_{l \in \mathcal{L}_{nm}} Q_{Lnm} = Q_{dm} \qquad \forall m \in \mathcal{M}
\end{equation}

In equation (20), the constraints on the power output of DGs are specified.
\begin{equation}
P_{\text{DG}_i}^{\min} \leq P_{\text{DG}_i} \leq P_{\text{DG}_i}^{\max} \quad \forall i \in \{1, 2, \ldots, N_{\text{DG}}\}
\end{equation}

The voltage magnitude at each bus is constrained within specified upper and lower bounds, as given in equation (21).

\begin{equation}
V_m^{\min} \leq V_m \leq V_m^{\max} \qquad \forall m \in \mathcal{M}
\end{equation}
\section{Proposed Methodology}
Heuristic-based approaches for the allocation of DG units in RDN are widely explored in recent literature. These approaches can provide approximate solutions with limited computational burden but often overlook critical operational constraints of network, such as line flow limits, equipment ratings, permissible bus voltage limits. Ignoring these constraints can lead to significant stress on the system and potential operational issues. Although a fully exhaustive OPF-based approach for DG allocation satisfies all network constraints, it imposes a high computational burden. To address these drawbacks, an efficient two-stage methodology has been proposed. Fig.1 shows the flowchart for proposed two-stage methodology.
\begin{figure}[H]
    \centering
    \includegraphics[width=0.75\linewidth]{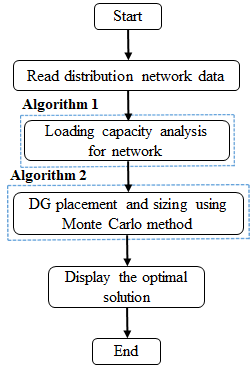}
    \caption{Flowchart of the proposed methodology}
\end{figure}
An iterative approach is used in the first stage to determine the additional active loading capacity of the network and of each individual bus.
This analysis determines the network’s additional active loading limit and identifies the buses with high active loading capacity, which are considered as the candidate nodes for DG placement. The steps involved for determining the additional loading capacity is provided in Algorithm 1. 
\begin{algorithm}
\caption{Loadability Capacity Analysis for Network}
\begin{algorithmic}
\State \textbf{Initialize} base power flow using MATPOWER
\State \textbf{Identify} load buses in the system data
\State \textbf{Set} system operational constraints
\For{each load bus in the system}
    \State \textbf{Initialize} $\lambda = 1$
    \State \textbf{Set} $\lambda_{\text{max}} = \lambda$
    \State \textbf{Scale} active power load at current bus using $\lambda$
    \State \textbf{Create} a test case with updated load
    \While{power flow converges}
        \State \textbf{Check} operational constraints
        \If{all constraints are satisfied}
            \State \textbf{Update} $\lambda_{\text{max}}$
            \State \textbf{Increment} $\lambda \gets \lambda + \lambda_{\text{step}}$
        \Else
            \State \textbf{Break}
        \EndIf
    \EndWhile
    \State \textbf{Store} $\lambda_{\text{max}}$ for current bus
\EndFor
\State \textbf{Calculate} Additional load = $(\lambda - 1) \times$ base load at each bus
\State \textbf{Identify} top $N$ buses with highest loadability margin 
\end{algorithmic}
\end{algorithm}

In the second stage, Monte Carlo method is employed to determine the optimal locations and sizes of DG units, with the objective of minimizing voltage deviation and reducing active power losses. The Monte Carlo method is executed $T$ times, with the placement of DG units on buses selected via probabilistic sampling from the top $N$ buses, and assigned sizes within the specified penetration limits of the network’s loadability.
For each trial, the system is modified with selected DG configurations, and power flow analysis is performed. If the power flow converges and all bus voltage constraints are satisfied, objective function evaluated and the trial results are stored. The steps involved in assessing the optimal placement and sizing is provided in Algorithm 2. 

\begin{algorithm}
\caption{DG Placement and Sizing using Monte Carlo Method}
\begin{algorithmic}
\State \textbf{Initialize:} Number of DG units $N_{DG}$, bounded DG limits, bus voltage limits, and number of Monte Carlo trials $T$
\For{each trial $t = 1$ to $T$}
    \State \textbf{Select} buses for DG placement using probabilistic sampling from the top $N$ candidate/ranked buses
    \State \textbf{Assign} DG sizes within specified bounded penetration limits
    \State \textbf{Modify} system data with selected DG placement and sizing
    \State \textbf{Run} power flow for modified system
    \If{power flow converges}
        \State \textbf{Check} voltage constraints at all buses
        \If{all constraints are satisfied}
            \State \textbf{Compute} objective function (OF)
            \State \textbf{Store} trial result (placement, sizing, OF)
        \Else
            \State \textbf{Discard} trial
        \EndIf
    \Else
        \State \textbf{Discard} trial
    \EndIf
\EndFor
\State \textbf{From} stored trials, identify the one with:
\begin{itemize}
    \item Valid voltage limits at all buses
    \item Minimum objective function value
\end{itemize}
\State \textbf{Compute} and display the optimal solution for DG placement and sizing from the stored trial
\end{algorithmic}
\end{algorithm}

\section{Validation Using test system}
The IEEE 33-bus RDN is used to validate the proposed two-stage methodology. The corresponding single-line diagram of RDN is in Fig. 2. RDN has a total load demand of 3.72 MW of active power and 2.3 MVar of reactive power. Under the base operating condition, the network experiences losses of 0.211 MW (active) and 0.143 MVar (reactive). Moreover, 21 of the 33 buses have voltages lower than the specified voltage limit of 0.95 p.u. \cite{28}. All simulations were performed in MATLAB using MATPOWER on a laptop with an Intel(R) Core(TM) i7 processor running at 3.20 GHz and 16 GB of RAM.
\begin{figure} [H]
    \centering
    \includegraphics[width=1\linewidth]{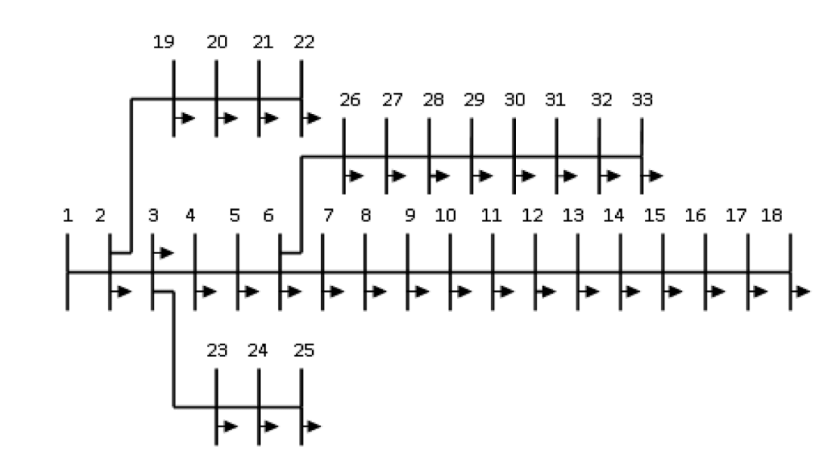}
    \caption{Single line diagram of test network [28].}
\end{figure}
\subsection{Results and Discussions}
 The increase in active loading capacity for each individual bus, within the limits of operational constraints, is shown in Fig. 3. It illustrates both the fixed load and the additional load that can be accommodated at each bus. Out of all the buses, Bus 2 can handle the maximum load increase, whereas Bus 18 allows the least. The maximum achievable load of 4.6 MW is obtained when simultaneous loading is permitted across all buses.
\begin{figure} [H]
    \centering
    \includegraphics[width=1\linewidth]{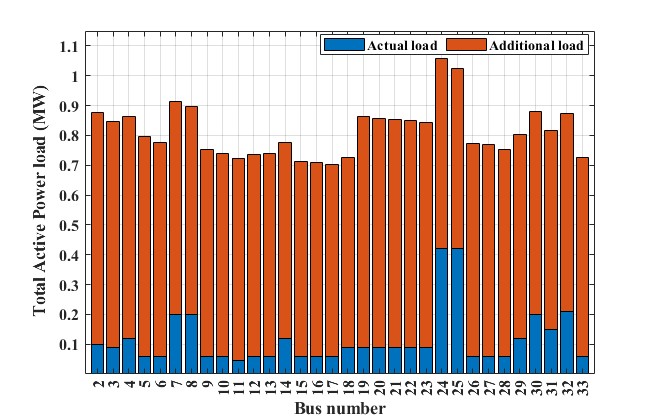}
    \caption{Fixed load and additional loading for each individual bus.}
\end{figure}
The proposed method is used to allocate PV-based DG units, which supply active power ($P_{\text{DG}}$).
Table I summarizes the optimal allocation results for one, two, and three DG units in RDN. For a single DG unit, bus 6 is identified as the optimal location with a capacity of 2.893 MW, lowering total active power losses from 0.211 MW to 0.1047 MW. The minimum bus voltage in this case is 0.9556 p.u., observed at bus 18.  For two DG units, the optimal placement is determined at buses 12 and 30, with corresponding capacities of 1.147 MW and 1.239 MW. This configuration results in a further reduction of the network’s active power losses to 0.0873 MW, while the minimum bus voltage improves to 0.9751 p.u., observed at bus 33. For three DG units, the optimal placement is determined at buses 13, 24, and 30, with corresponding capacities of 1.032 MW, 1.096 MW, and 1.062 MW, respectively. The lowest active power losses of 0.0735 MW are obtained, with a minimum bus voltage of 0.9556 p.u. observed at bus 18.

\begin{table}[ht]
\scriptsize
\caption{Results for Allocation of DG Units}
\resizebox{\columnwidth}{!}{
\begin{tabular}{|c|c|c|c|c|c|}
\hline
\shortstack{\rule{0pt}{2.5ex}\textbf{No. of DG }\\\textbf{Units}} & 
\shortstack{\rule{0pt}{2.5ex}\textbf{Optimal Loc-}\\\textbf{ation(s)}} & 
\shortstack{\rule{0pt}{2.5ex}\textbf{DG Size }\\\textbf{(MW)}} & 
\shortstack{\rule{0pt}{2.5ex}\textbf{Losses w/o}\\\textbf{DG}} & 
\shortstack{\rule{0pt}{2.5ex}\textbf{Losses with}\\\textbf{DGs}} & 
\shortstack{\rule{0pt}{2.5ex}\(\boldsymbol{V_{\min}}\)\textbf{(p.u.)}, \\\textbf{Bus}} \\
\hline
One DG Unit & \rule{0pt}{2.5ex} 6 & 2.893 & 0.211 & 0.1047 & 0.9556, 18 \\
\hline
\raisebox{2ex}{Two DG Units} & \shortstack[t]{\rule{0pt}{2.5ex}12\\30} & \shortstack[t]{1.147\\1.239} & \raisebox{2.5ex}{0.211} & \raisebox{2.5ex}{0.0873} & \raisebox{2.5ex}{0.9751, 33} \\
\hline
\raisebox{5ex}{Three DG Units} & \shortstack[t]{\rule{0pt}{2.5ex}13\\24\\30} & \shortstack[t]{1.032\\1.096\\1.062} &\raisebox{5ex}{0.211} & \raisebox{5ex}{0.0735} & \raisebox{5ex}{0.9721, 33} \\
\hline
\end{tabular}}
\end{table}

Fig. 4 illustrates the voltage magnitude profile of the RDN under four scenarios: without DG, with single DG, with two DGs, and with three DGs. Under the base operating condition or without any DG unit, the network exhibits significant voltage drops, with  minimum voltage around 0.91 p.u. at farthest buses.
With a single DG unit, the voltage profile shows noticeable improvement, with minimum bus voltage rises above 0.95 p.u. and 17 buses having voltage magnitudes greater than 0.98 p.u. With two DG units, the voltage profile across the buses remains above 0.97 p.u., and 23 buses have voltage magnitudes above 0.98 p.u.
With three DG units, the system achieves the best performance, with voltage magnitudes maintained above 0.97 p.u. across all buses and 28 buses have voltage magnitude more than 0.98 p.u.
\begin{figure}[H]
    \centering
    \includegraphics[width=1.02\columnwidth]{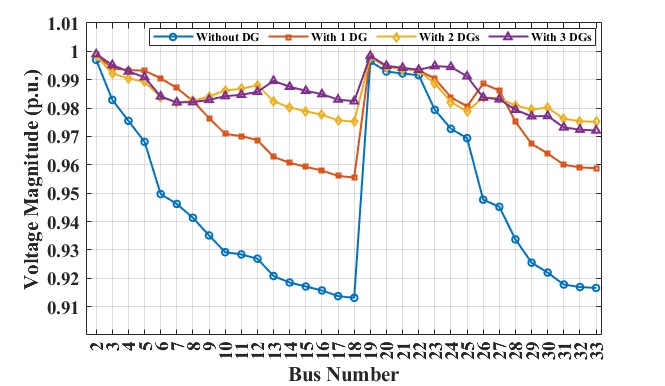}
    \caption{RDN voltage magnitude profiles for different DG scenarios.}
\end{figure}
\subsection{Comparative analysis of the proposed and existing methods}
 Tables II, III, and IV presents a comparison of the results obtained from  proposed two-stage methodology  for allocation of DG units, with those of several existing methodologies such as improved analytical technique (IAT) [29], genetic algorithm (GA) [30], salp swarm algorithm (SSA) [31], multi-objective whale optimization algorithm (WOA) [32],  efficient analytical (EA) [33], coot bird optimization (CBOM) [34], and improved HHO using PSO [35]. For the case of a single DG unit, all methodologies identify bus 6 as the optimal location, but the proposed method allows the largest DG size of 2.893 MW, achieves 50.37\% reduction in active power losses which is higher than most existing methods, and obtain $\text V_{\text{min}}$ of 0.9556 p.u., which is better than $\text V_{\text{min}}$ obtained from all other methods, as summarized in Table II.

\begin{table}[H]
\scriptsize 
\setlength{\tabcolsep}{6pt} 
\renewcommand{\arraystretch}{1.5} 
\caption{Single DG Allocation in RDN: A Comparative Study of Different Methods}
\begin{tabular}{|c|c|c|c|c|c|}
\hline
\shortstack{\rule{0pt}{2.5ex}\textbf{}\\\textbf{DG}\\\textbf{units}} &
\shortstack{\rule{0pt}{2.5ex}\textbf{Different}\\\textbf{Methods}} &
\multicolumn{4}{c|}{\rule{0pt}{2.5ex}\textbf{For IEEE 33 Bus RDN}} \\
\cline{3-6}
& & \raisebox{2.5ex}{\textbf{Bus No.}}& 
\shortstack{\textbf{DG Size}\\\textbf{(MW)}} & 
\shortstack{\textbf{ }\\\textbf{Loss Red-}\\\textbf{uction (\%)}}& 
\shortstack{\(\boldsymbol{V_{\min}}\) \\\textbf{(p.u.)}}\\
\hline            

\multirow{7}{*}{1} 
 & IAT\cite{29} & 6 & 2.431 & 47.31 & 0.9476 \\
\cline{2-6}
 & GA\cite{30} & 6 & 2.500 & 46.96 & 0.9500 \\
\cline{2-6}
 & SSA\cite{31} & 6 & 2.490 & 47.31 & 0.9401 \\
\cline{2-6}
 & WOA\cite{32} & 6 & 2.590 & 47.39 & 0.9425 \\
\cline{2-6}
 & EA\cite{33} & 6 & 2.530 & 47.39 & 0.9504 \\
\cline{2-6}
 & CBOM\cite{34} & 6 & 2.575 & 50.72 & 0.9511 \\
\cline{2-6}
 & HHOPSO\cite{35} & 6 & 2.574 & 50.73 & 0.9510 \\
\cline{2-6}
 & \textbf{Proposed Method} & \textbf{6} & \textbf{2.893} & \textbf{50.37} & \textbf{0.9556} \\
\hline
\end{tabular}
\end{table}

For the case of two DG units, the proposed method selects buses 12 and 13 as optimal locations. The total DGs capacity is 2.380 MW, which is larger than that of other methods, while achieving a comparable active power losses reduction of 58.62\% and $\text V_{\text{min}}$  of 0.9751 p.u., which is better than $\text V_{\text{min}}$ obtained from all other methods, as summarized in Table III.

\begin{table}[H]
\scriptsize 
\setlength{\tabcolsep}{3pt} 
\renewcommand{\arraystretch}{1.5} 
\caption{Two DG Allocation in RDN: A Comparative Study of Different Methods}
\begin{tabular}{|c|c|c|c|c|c|c|}
\hline
\shortstack{\rule{0pt}{2.5ex}\textbf{}\\\textbf{DG}\\\textbf{units}} & 
\shortstack{\textbf{Different}\\\textbf{Methods}} & 
\multicolumn{5}{c|}{\textbf{For IEEE 33 Bus RDN}} \\
\cline{3-7}
& & 
\raisebox{2.5ex}{\textbf{Bus No.}} & 
\shortstack{\textbf{DG Size}\\\textbf{(MW)}} & 
\shortstack{\textbf{ }\\\textbf{Loss Red-}\\\textbf{uction (\%)}}&   
\shortstack{\textbf{Total DG}\\\textbf{Size (MW)}} & 
\shortstack{\(\boldsymbol{V_{\min}}\) \\ \textbf{(p.u.)}}\\
\hline
\multirow{4}{*}{2}
 & \shortstack{IAT\cite{29}}& - & - & 55.62 & - & - \\
 \cline{2-7} & \raisebox{2.5ex}{GA\cite{30}} & \shortstack[t]{\rule{0pt}{2.5ex}14\\30} & \shortstack{0.750\\1.250} & \raisebox{2.5ex}{58.44} & \raisebox{2.5ex}{2.000} & \raisebox{2.5ex}{0.9701} \\
\cline{2-7}
 & \raisebox{2.5ex}{SSA\cite{31}} & \shortstack[t]{\rule{0pt}{2.5ex}13\\30} & \shortstack[t]{\rule{0pt}{2.5ex}0.832\\1.110} & \raisebox{2.5ex}{58.64} & \raisebox{2.5ex}{1.970} & \raisebox{2.5ex}{0.9667} \\
\cline{2-7}
 & WOA\cite{32} & - & - & - & - & - \\
\cline{2-7}
 & \raisebox{2.5ex}{EA\cite{33}} & \shortstack[t]{\rule{0pt}{2.5ex}13\\30} & \shortstack[t]{\rule{0pt}{2.5ex}0.844\\1.140} & \raisebox{2.5ex}{58.50} & \raisebox{2.5ex}{1.958} & \raisebox{2.5ex}{0.9682} \\
\cline{2-7}
 & \raisebox{2.5ex}{CBOM\cite{34}} & \shortstack[t]{\rule{0pt}{2.5ex}13\\30} & \shortstack[t]{\rule{0pt}{2.5ex}0.852\\1.157} & \raisebox{2.5ex}{58.69} & \raisebox{2.5ex}{2.009} & \raisebox{2.5ex}{0.9684} \\
\cline{2-7}
 & \raisebox{2.5ex}{HHOPSO\cite{35}} & \shortstack[t]{\rule{0pt}{2.5ex}13\\30} & \shortstack[t]{\rule{0pt}{2.5ex}0.846\\1.158} & \raisebox{2.5ex}{59.09} & \raisebox{2.5ex}{2.004} & \raisebox{2.5ex}{0.9680} \\
\cline{2-7}
 & \raisebox{2.5ex}{\textbf{Proposed Method}} & \shortstack[t]{\rule{0pt}{2.5ex}\textbf{12}\\\textbf{30}} & \shortstack[t]{\rule{0pt}{2.5ex}\textbf{1.147}\\\textbf{1.239}} & \raisebox{2.5ex}{\textbf{58.62}} & \raisebox{2.5ex}{\textbf{2.380}} & \raisebox{2.5ex}{\textbf{0.9751}} \\
\hline
\end{tabular}
\end{table}

 For the case of three DG units, the optimal buses are 13, 24 and 30. The total DGs capacity is 3.190 MW, which is larger than that of other methods, while achieving a 65.16\% reduction in active power losses, which is only about 1 kW higher than that obtained by some existing methods and $\text V_{\text{min}}$ of 0.9721 p.u., which is better $\text V_{\text{min}}$ obtained from all other methods, as summarized in Table IV. The increased capacity of DGs results in significantly improved voltage magnitude profile throughout the RDN, ensuring that voltages remain within permissible operating limits. Figs. 5–7 illustrate the improved bus voltage profiles of the RDN for three cases: with one, two, and three DG units, respectively. As evident from the figures, the voltage profile improves progressively with an increasing number of DG units across all methods. However, compared to existing methods, the proposed method demonstrates an improved voltage profile in all three cases by maintaining higher bus voltage magnitudes throughout RDN. 
\begin{table}[H]
\centering
\scriptsize 
\setlength{\tabcolsep}{3.1pt} 
\renewcommand{\arraystretch}{1.2} 
\caption{Three DG Allocation in RDN: A Comparative Study of Different Methods}
\begin{tabular}{|c|c|c|c|c|c|c|}
\hline
\shortstack{\rule{0pt}{2.5ex}\textbf{}\\\textbf{DG}\\\textbf{units}}& 
\shortstack{\textbf{Different}\\\textbf{Methods}} & 
\multicolumn{5}{c|}{\textbf{For IEEE 33 Bus RDN}} \\
\cline{3-7}
& & 
\raisebox{2.5ex}{\textbf{Bus No.}} & 
\shortstack{\textbf{DG Size}\\\textbf{(MW)}} & 
\shortstack{\textbf{ }\\\textbf{Loss Red-}\\\textbf{uction(\%)}}& 
\shortstack{\textbf{Total DG}\\\textbf{Size (MW)}} & 
\shortstack{\(\boldsymbol{V_{\min}}\) \\ \textbf{(p.u.)}}\\
\hline
\multirow{8}{*}{3}
& \raisebox{5ex}{IAT \cite{29}}
  & \shortstack[t]{\rule{0pt}{2.5ex}6\\18\\32} 
  & \shortstack[t]{\rule{0pt}{2.5ex}1.312\\0.462\\0.657} 
  & \raisebox{5ex}{61.48} 
  & \raisebox{5ex}{2.431} 
  & \raisebox{5ex}{0.9701} \\
\cline{2-7}
& \raisebox{5ex}{GA \cite{30}} 
  & \shortstack[t]{\rule{0pt}{2.5ex}14\\24\\29} 
  & \shortstack[t]{\rule{0pt}{2.5ex}0.750\\1.250\\1.000} 
  & \raisebox{5ex}{64.95} 
  & \raisebox{5ex}{3.000} 
  & \raisebox{5ex}{0.9639} \\
\cline{2-7}
& \raisebox{5ex}{SSA \cite{31}} 
  & \shortstack[t]{\rule{0pt}{2.5ex}13\\24\\30} 
  & \shortstack[t]{\rule{0pt}{2.5ex}0.790\\1.070\\1.012} 
  & \raisebox{5ex}{65.45} 
  & \raisebox{5ex}{2.872} 
  & \raisebox{5ex}{0.9670} \\
\cline{2-7}
& \raisebox{5ex}{WOA \cite{32}} 
  & \shortstack[t]{\rule{0pt}{2.5ex}13\\24\\30} 
  & \shortstack[t]{\rule{0pt}{2.5ex}0.801\\1.091\\1.053} 
  & \raisebox{5ex}{65.50} 
  & \raisebox{5ex}{2.945} 
  & \raisebox{5ex}{0.9687} \\
\cline{2-7}
& \raisebox{5ex}{EA \cite{33}} 
  & \shortstack[t]{\rule{0pt}{2.5ex}13\\24\\30} 
  & \shortstack[t]{\rule{0pt}{2.5ex}0.798\\1.099\\1.050} 
  & \raisebox{5ex}{65.36} 
  & \raisebox{5ex}{2.947} 
  & \raisebox{5ex}{0.9683} \\
\cline{2-7}
& \raisebox{5ex}{CBOM \cite{34}} 
  & \shortstack[t]{\rule{0pt}{2.5ex}13\\24\\30} 
  & \shortstack[t]{\rule{0pt}{2.5ex}0.802\\1.091\\1.054} 
  & \raisebox{5ex}{65.50} 
  & \raisebox{5ex}{2.947} 
  & \raisebox{5ex}{0.9687} \\
\cline{2-7}
& \raisebox{5 ex}{HHOPSO \cite{35}} 
  & \shortstack[t]{\rule{0pt}{2.5ex}14\\24\\30} 
  & \shortstack[t]{\rule{0pt}{2.5ex}0.761\\1.094\\1.068} 
  & \raisebox{5ex}{66.14} 
  & \raisebox{5ex}{2.926} 
  & \raisebox{5ex}{0.9687} \\
\cline{2-7}
& \raisebox{5 ex}{\textbf{Proposed Method}} 
  & \shortstack[t]{\rule{0pt}{2.5ex}\textbf{13}\\\textbf{24}\\\textbf{30}} 
  & \shortstack[t]{\rule{0pt}{2.5ex}\textbf{1.032}\\\textbf{1.096}\\\textbf{1.062}} 
  & \raisebox{5ex}{\textbf{65.16}} 
  & \raisebox{5ex}{\textbf{3.190}} 
  & \raisebox{5ex}{\textbf{0.9721}} \\
\hline

\end{tabular}
\end{table}

\begin{figure}[H]
    \centering
    \includegraphics[width=\columnwidth, height=4.9cm]{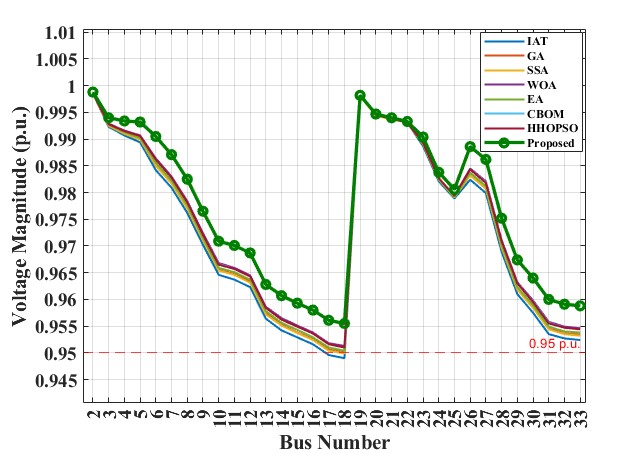}
    \caption{Improved bus voltage profiles of the RDN with one DG unit.}
\end{figure}

\begin{figure}[H]
    \centering
    \includegraphics[width=\columnwidth, height=5cm]{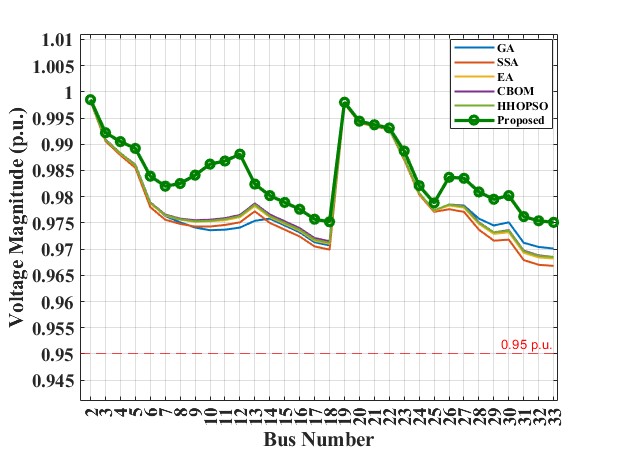}
    \caption{Improved bus voltage profiles of the RDN with two DG units.}
\end{figure}

\begin{figure}[H]
    \centering
    \includegraphics[width=\columnwidth, height=5.1cm]{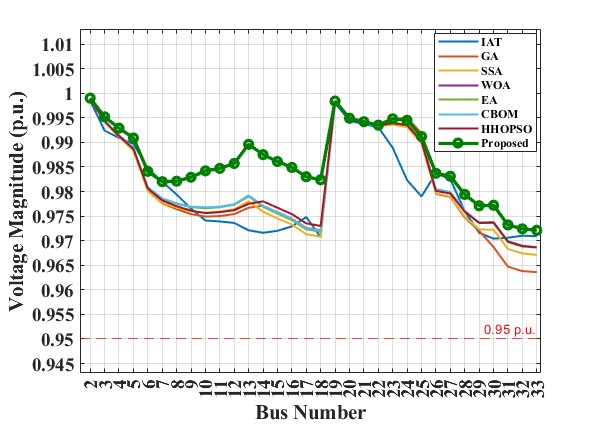}
    \caption{Improved bus voltage profiles of the RDN with three DG units.}
\end{figure}

\section{Conclusion}
This paper presents an efficient two-stage methodology for the optimal allocation of multiple PV-based DG units in a radial distribution network, while accounting for the network’s loading capacity. An iterative technique is used in the first stage to evaluate the additional active loading capacity of the network and each individual bus, identifying candidate locations for DG placement. The Monte Carlo method is used in the second stage to determine the optimal sizes and locations of the DG units, aiming to minimize voltage deviations and reduction in active power losses in the network. The proposed two-stage methodology, when validated on IEEE 33-bus RDN, demonstrates notable improvements in overall network performance. Specifically, the optimal allocation of one, two, and three DG units yields active power losses reductions of 50.37\%, 58.62\%, and 65.16\%, respectively, while simultaneously improving bus voltage magnitudes throughout the RDN.

The comparative analysis of the proposed methodology with other existing methods demonstrates its merits, as it identifies suitable DG locations, accommodates larger DG capacities, and achieves significant reductions in active power losses, and notable voltage profiles improvements. These advantages highlight the practical applicability of the method for integrating DGs into modern distribution systems. Future work aims to determine overall loadability limits  of the network and of each individual bus, as well as to explore the integration of other types of DG units.


%

\ifCLASSOPTIONcaptionsoff
  \newpage
\fi



%

\bibliographystyle{ieeetr}     
\bibliography{references}      

%




\end{document}